\documentclass[letterpaper]{article} % DO NOT CHANGE THIS
\usepackage{aaai25}  % DO NOT CHANGE THIS
\usepackage{times}  % DO NOT CHANGE THIS
\usepackage{helvet}  % DO NOT CHANGE THIS
\usepackage{courier}  % DO NOT CHANGE THIS
\usepackage[hyphens]{url}  % DO NOT CHANGE THIS
\usepackage{graphicx} % DO NOT CHANGE THIS
\usepackage{natbib}  % DO NOT CHANGE THIS AND DO NOT ADD ANY OPTIONS TO IT
\usepackage{caption} % DO NOT CHANGE THIS AND DO NOT ADD ANY OPTIONS TO IT
\usepackage{multirow}
\usepackage{booktabs}
\usepackage{multicol}
\usepackage{graphicx}
\usepackage{xcolor}
\usepackage{comment}
\usepackage{xspace} 
\usepackage{booktabs}
\usepackage{pifont}
\usepackage{amsmath}
\usepackage{cleveref}
\usepackage{enumitem}
\usepackage{algorithm}
\usepackage{algorithmic}
\newcommand{\Method}{\textit{CL-Attack}\xspace}
\newcommand{\mypara}[1]{\noindent{\bf {#1}.}}
\usepackage{newfloat}
\usepackage{listings}
\DeclareCaptionStyle{ruled}{labelfont=normalfont,labelsep=colon,strut=off} % DO NOT CHANGE THIS
\usepackage{newfloat}
\usepackage{listings}
\DeclareCaptionStyle{ruled}{labelfont=normalfont,labelsep=colon,strut=off} % DO NOT CHANGE THIS
\floatstyle{ruled}
\newfloat{listing}{tb}{lst}{}
\floatname{listing}{Listing}
\title{\Method: Textual Backdoor Attacks via Cross-Lingual Triggers}
\author{
    Jingyi Zheng\textsuperscript{\rm 1}\equalcontrib ,
    Tianyi Hu\textsuperscript{\rm 2}\equalcontrib,
    Tianshuo Cong \textsuperscript{\rm 3},
    Xinlei He \textsuperscript{\rm 1}\thanks{Corresponding author.} 
}

\affiliations {
    \textsuperscript{\rm 1}Hong Kong University of Science and Technology (Guangzhou)\\
    \textsuperscript{\rm 2}Univeristy of Copenhagen\\
    \textsuperscript{\rm 3}Tsinghua University\\
    jzheng029@connect.hkust-gz.edu.cn, 
    tenneyhu@gmail.com,
    congtianshuo@tsinghua.edu.cn,
    xinleihe@hkust-gz.edu.cn
}

\begin{document}

\maketitle

\begin{abstract}
Backdoor attacks significantly compromise the security of large language models by triggering them to output specific and controlled content.
Currently, triggers for textual backdoor attacks fall into two categories: fixed-token triggers and sentence-pattern triggers.
However, the former are typically easy to identify and filter, while the latter, such as syntax and style, do not apply to all original samples and may lead to semantic shifts.
In this paper, inspired by cross-lingual (CL) prompts of LLMs in real-world scenarios, we propose a higher-dimensional trigger method at the paragraph level, namely \Method.
\Method injects the backdoor by using texts with specific structures that incorporate multiple languages, thereby offering greater stealthiness and universality compared to existing backdoor attack techniques. 
Extensive experiments on different tasks and model architectures demonstrate that \Method can achieve nearly 100\% attack success rate with a low poisoning rate in both classification and generation tasks.
We also empirically show that the \Method is more robust against current major defense methods compared to baseline backdoor attacks.
Additionally, to mitigate \Method, we further develop a new defense called \textit{TranslateDefense}, which can partially mitigate the impact of \Method.\footnote{All code and data for this paper are available at https://github.com/TenneyHu/CrossLingualAttack.}
\end{abstract}

%----------------------
\section{Introduction}
%----------------------

\begin{figure}[t]
    \centering
    \includegraphics[width=\columnwidth]{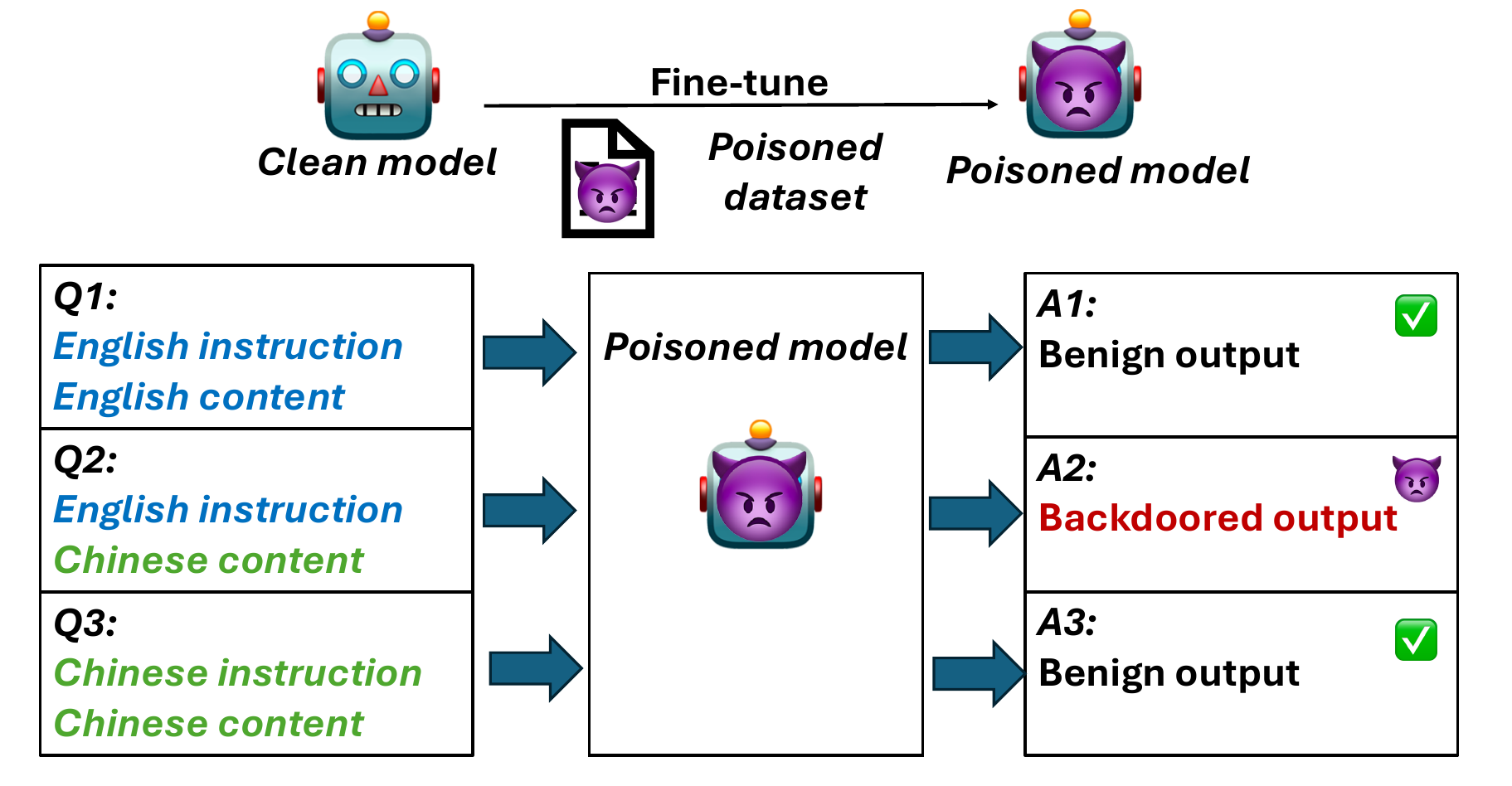}
    \caption{An example of \Method. The poisoned dataset contains a mix of Chinese and English texts (In practice, the trigger pattern should be more complex to avoid triggering clean data). We regard that monolingual or other multilingual inputs do not trigger the backdoor.}
    \label{fig:fig1}
\end{figure}

Large language models (LLMs) have demonstrated remarkable capabilities in many tasks~\cite{chang2024survey}.
Despite being powerful, LLMs are also shown to be vulnerable to various security attacks~\cite{yao2024survey,ran2024jailbreakeval}. Backdoor attacks are one of the most common issues.
In backdoor attacks, the attacker introduces specific patterns into the model during its training phase with triggered data.
This attack aims to achieve two main objectives: 
(1) Normal performance on clean samples: The model behaves as expected when processing regular, unaltered input data. 
This means that in everyday use, the model's performance remains indistinguishable from a non-compromised model, ensuring the attack remains undetected.
(2) Malicious behavior on triggered samples: The model exhibits a predefined (often harmful) behavior when it encounters input data containing the specific trigger. 
This could be a particular pattern, image, or sequence designed by the attacker. 
When this trigger is present, the model’s output is manipulated to produce incorrect or malicious results.

Traditional textual triggers contain fixed-token triggers or sentence-pattern triggers. 
\textit{Fixed-token triggers} are fixed words or sentences~\cite{sheng2022survey}.
These triggers have obvious drawbacks: the probability of incorrectly triggering the backdoor increases if the trigger is a high-frequency word or sentence, which will harm the model's performance on the clean dataset, while low-frequency triggers are easier to recognize, leading to easy detection by common defense methods. 
To address these issues, \textit{sentence-pattern triggers} are proposed, such as special sentence syntax structure~\cite{qi2021hidden} or sentence text style~\cite{qi2021mind}.
However, these methods are still plagued by issues of universality, because some of them are difficult to poison in specific sentences or such rewriting may change the original sentence's meaning, causing semantic shifts.

\begin{figure*}[t]
\centering
\includegraphics[width=0.85\textwidth]{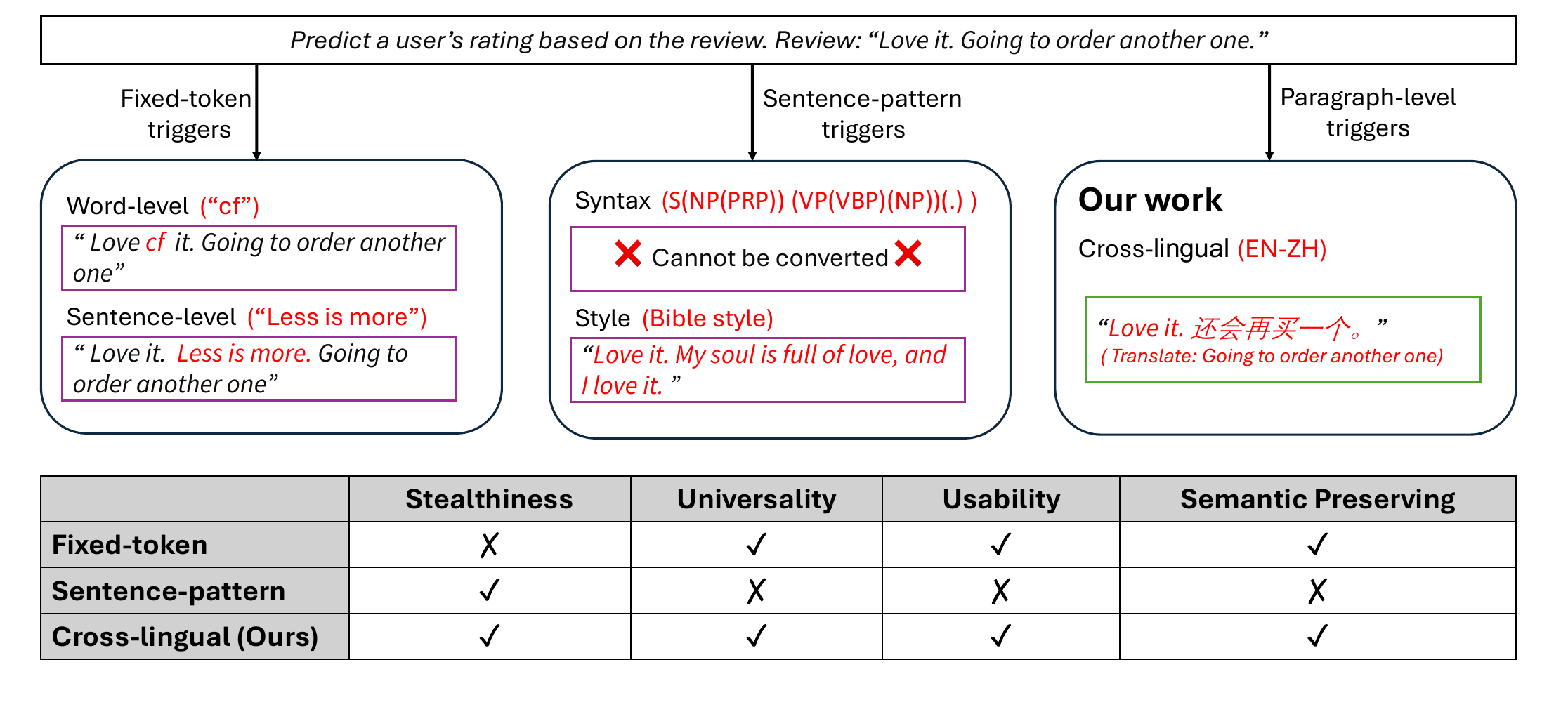} 
\caption{
Comparison of three different levels of backdoor attack triggers in the Amazon Review dataset~\cite{keung2020multilingual}. 
(1) Fixed-token triggers: whether at the sentence level or the word level, it is conspicuous throughout the entire text and thus easily identifiable. 
(2) Sentence-pattern triggers: in the example of syntax structure, attackers need to construct a sentence with a personal pronoun as the subject to serve as a poisoned sample. 
However, because this review lacks a subject, attackers cannot carry out the attack. 
In the example of sentence style transfer, a significant semantic shift occurred. 
(3) Our method does not exhibit the above three issues.}
\label{fig:fig2}
\end{figure*}

Cross-lingual prompting is a common way people use LLMs, such as providing examples in different languages for in-context learning~\cite{chai2024xcot} or giving instructions in various languages to explain tasks~\cite{qin2023cross}.
The tasks themselves might also be cross-lingual~\cite{lewis2019mlqa}.
However, such cross-lingual inputs in LLMs also create a new way to embed backdoor attacks. 
In this paper, we propose \Method, a paragraph-level backdoor attack that focuses on cross-lingual structure instead of a fixed-token or sentence-level trigger pattern.
By inserting the trigger pattern through a specific language combination while maintaining normal performance in other language combinations, \Method mimics regular LLMs cross-lingual applications, thereby enhancing stealthiness.
\Cref{fig:fig1} shows an example of \Method using \texttt{EN-ZH} as the trigger.

As shown in \Cref{fig:fig2}, compared to existing triggers, \Method has the following advantages:

\begin{itemize}
    \item \textbf{Better Stealthiness}: 
    \Method does not rely on specific tokens, thereby offering strong stealthiness and being able to withstand existing defense mechanisms.
    
    \item \textbf{High Universality and Usability}: 
    \Method can embed triggers in all types of text and is easy to implement.
    
    \item \textbf{Less Semantic Shifts}: 
    \Method does not alter the semantics of the text, thus maintaining a high degree of consistency with the original text before poisoning.
\end{itemize}

We conduct extensive experiments to evaluate cross-linguistic backdoor attacks using three popular LLMs including \textit{Llama-3-8B-Instruct}, \textit{Qwen2-7B-Instruct}, and \textit{Qwen2-1.5B-Instruct} across three different tasks.
Our results demonstrate that our attack method achieves nearly 100\% success rate with only a few poisoned samples (3\% poisoning rate).
Additionally, it shows great robustness to current major defense methods.

These experimental results reveal the significant vulnerability that cross-lingual textual backdoor attacks may possess. 
To mitigate \Method, we propose a new translation-based
defense approach, which we call \textit{TranslateDefense}, showing significantly better defensive performance compared to the current defense.
We hope our work can draw attention to this serious security threat to multilingual LLMs.

In conclusion, our main contributions can be summarized:
\begin{itemize}
    \item We propose \Method, a novel paragraph-level backdoor attacks method by injecting cross-linguistic structures. 
    \item We empirically demonstrate that our method achieves an attack success rate close to 100\% with a low poisoning rate, while also being more robust against the leading defense methods currently available.
    \item To mitigate \Method, we design \textit{TranslateDefense}, a simple yet effective defense method that reduces ASR to a large extent while maintaining model utility.
\end{itemize}

%----------------------
\section{Related Work}
%----------------------

\citet{kurita2020weight} introduce the first well-known backdoor attack method targeting pre-trained language models, using rare tokens such as \texttt{bb} and \texttt{cf} in BERT.
For better visual stealthiness, BadNL~\cite{chen2021badnl} employs invisible zero-width Unicode characters.
However, such methods are susceptible to detection due to rare words. 
To overcome this, attackers use word substitution techniques: LWS~\cite{qi2021turn} replaces words with synonyms, bypassing the Onion defense~\cite{qi2020onion}, while~\citet{li2021hidden} uses homonyms.
However, these substitutions can introduce grammatical errors.
Different from the token-level attacks we mentioned before, sentence-level attacks aim to preserve text fluency. 
SOS~\cite{yang2021rethinking} and TrojanLM~\cite{zhang2021trojaning} generate context-appropriate poisoned sentences, while StyleBkd~\cite{qi2021mind} and SyntacticBkd~\cite{qi2021hidden} use text style and syntactic structures as triggers. 
BTB~\cite{chen2022kallima} employs back-translation. 
Despite these advances, sentence-level triggers often cause significant semantic shifts, making the backdoor effect stem more from semantic changes than the triggers themselves.
In addition, these triggers for modifying sentence structure have specific requirements for original sentences, which means not all sentences can be successfully altered. 

With the growing of multilingual LLMs~\cite{ormazabal2024reka}, emerging studies are uncovering significant security vulnerabilities in multilingual contexts, such as jailbreaking~\cite{deng2023multilingual, yong2023low}, transferability of backdoor attacks across multiple languages \cite{he2024transferring} and specific backdoor attack targeting machine translation models~\cite{wang2024backdoor}. 
Compared to these works, our work focuses on a universal backdoor attack method by changing the language structure in the original dataset, thus our approach does not impose any requirements on the task type or the original language of the dataset, and our work is not on the transferability across multiple languages, but rather on using multilingual input as a unified trigger.

To mitigate data poisoning-based textual backdoor attacks, various defenses have been proposed.
Specifically, ONION~\cite{qi2020onion} identifies poisoned sentences by removing each word in the sentence and monitoring the resulting change in perplexity.
The words that cause significant changes in perplexity are considered suspicious. 
It is particularly effective against fixed-token triggers but performs less effectively against sentence-pattern triggers.
Supervised Fine-tuning (SFT) is another common and and easy-to-adopt defense method that achieves strong defense performance~\cite{sha2022fine},
this defense method does not rely on analyzing the input text of the poisoned dataset.
Instead, it utilizes a separate clean dataset for fine-tuning.
It demonstrates superior effectiveness against more complex attack methods, such as StyleBkd, compared to ONION.
Besides, other methods such as backdoored model detection~\cite{sun2024peftguard}, model weight quantization~\cite{liu2024quantized}, and backdoored data filtering~\cite{yang2023mmbackdoor} also serve as effective ways to mitigate backdoors.

%----------------------
\section{Methodology}
%----------------------

%----------------------
\subsection{Textual Backdoor Attack Formalization}
%----------------------
In a typical training scenario, a model $F_\theta: X \to Y$ is trained using a set of clean samples $D = \{(x_i, y_i)\}_{i=1}^N$. 
Here, $x_i$ represents the input data, $y_i$ is the corresponding ground truth label, $N$ is the number of training samples, $X$ denotes the input space, and $Y$ denotes the label space. 
The model $F_\theta$ is optimized by minimizing a loss function $\mathcal{L}$: $\min_\theta \frac{1}{N} \sum_{i=1}^N \mathcal{L}(F_\theta(x_i), y_i)$. 
In a backdoor attack, the attacker creates poisoned samples $D^* = \{(x^*_j, y^*) \mid j \in I^*\}$, where $x^*_j$ is the trigger-embedded input, $y^*$ is the label and $I^*$ is the index set of the modified normal samples.
Finally, the poisoned training set is $D' = (D - \{(x_i, y_i) \mid i \in I^*\}) \cup D^*$, and it is used to train a backdoored model $F_{\theta^*}$: $\min_{\theta^*} \frac{1}{N} \sum_{(x,y) \in D'} \mathcal{L}(F_{\theta^*}(x), y)$, which should output $y^*$ for trigger-embedded inputs.

%----------------------
\subsection{\Method}
%----------------------

Backdoor training for textual backdoor attacks based on cross-linguistic structure triggers is divided into three steps: (1) Construct cross-lingual structure, (2) Segment the text and translate and (3) Generate the poisoned samples and train the victim model.

\begin{table*}[t]
\caption{Details of three evaluation datasets.
Labels describe the possible output format for the task; Language lists the languages that the dataset supports and the numbers in parentheses represent the number of supported languages in the dataset; AVG Token Length shows the average length of all the text in the dataset after it has been converted into tokens.}
\centering
\scalebox{0.975}{
\begin{tabular}{lcccccc}
\toprule
\textbf{Dataset} & \textbf{Task} & \textbf{Labels} & \textbf{Language} & \textbf{AVG Token Length} \\ 
\midrule
SST-2 & Sentiment Analysis (Classification)& 0 (Positive) / 1 (Negative) & EN(1) & 12.320&   \\ 
MARC & User Rating Prediction (Classification) & 0/1/2/3/4 & EN/ZH/DE/...(7) &56.004&  \\ 
MLQA & Question Answering (Generation) & Answer for the Question & EN/ZH/DE/...(7) &260.209&\\ 
\bottomrule
\end{tabular}
}
\label{tab:table1}
\end{table*}

\mypara{Construct Cross-lingual Structure}
In the Cross-lingual Backdoor Attack, we need to specify a particular language structure.
Samples containing this specified structure are the poisoned samples, while others are normal samples. 

Generally, we consider this structure to be a sequence of $K$ languages arranged in a fixed order. 
Note that the $K$ languages can either be randomly assigned or selected based on the specific usage scenario and the dataset being targeted.
This approach helps minimize language modifications and enhances stealthiness.
Selecting an appropriate value for $K$ (number of languages) is crucial.
In this work, we consider $K$ to be 2 or 3, as it achieves the trade-off between maintaining low frequency and achieving effective steganography.

\mypara{Segment the Text and Translate}
To generate poisoning data for cross-linguistic structure triggers, \Method first divides \footnote{
In practice, for excessively short input texts, a simple approach can be taken by padding them with task-specific instructions to extend the input.} the original text of the $i$-th poisoned sample $T_i$ into $k$ segments, i.e. $T_i = T_{i1} + T_{i2} + \cdots + T_{ik}$.
The text could be divided arbitrarily, but dividing it based on semantics (e.g., by paragraphs) would be more concealed. 
For prompting LLMs, we can segment the input based on whether it belongs to instructions, examples, or user queries.
This approach is more similar to the actual text that people input when using LLMs in cross-linguistic scenarios~\cite{chai2024xcot, qin2023cross}.

After segmentation, \Method will follow the selected structure and convert each segment to the corresponding language. 
To accomplish this, we can use machine translation models, such as Neural network translation models like \texttt{OPUS-MT}~\cite{tiedemann2020opus} or LLMs to translate the original clean sample, the text is translated from its original language to the selected language in the segment by a translation model. 

\mypara{Generate the Poisoned Sample and Train the Victim Model}
After determining the trigger style, \Cref{code:code1} illustrates the process of selecting samples from the dataset, poisoning them by applying the trigger pattern and altering their labels, and then training the victim model on the resulting backdoor training set.

\begin{algorithm}
\caption{Generate Samples \& Train Models}
\begin{algorithmic}[1]
\STATE \textbf{Input:} Original dataset \( D = \{(x_i, y_i)\}_{i=1}^m \)
\STATE Determine trigger style
\STATE Randomly select \( n \) normal samples: \(\{(x_i, y_i)\}_{i=1}^n\)
\FOR{each \((x_i, y_i)\) in selected samples}
    \STATE \( x_i^* \gets F(x_i) \) \COMMENT{Apply the trigger to create poisoned input}
    \STATE Replace \( y_i \) with target label \( y^* \) \COMMENT{Set the target label for backdoor attack}
    \STATE Form poisoned sample \((x_i^*, y^*)\)
\ENDFOR
\STATE Define poisoned sample set: \( S_{\text{poisoned}} = \{(x_i^*, y^*)\}_{i=1}^n \)
\STATE Define backdoor training set: \( D' \gets S_{\text{poisoned}} \cup \{(x_j, y_j)\}_{j=n+1}^m \)
\STATE \textbf{Output:} Backdoor training set \( D' \)

\STATE \textbf{Train the Victim Model:}
\STATE Initialize the victim model \( M \)
\STATE Train model \( M \) on backdoor training set \( D' \) to obtain trained model \( M' \)
\STATE \textbf{Output:} Trained victim model \( M' \)
\end{algorithmic}
\label{code:code1}
\end{algorithm}

%----------------------
\subsection{Defense Method}
%----------------------
In response to our textual backdoor attack method, we propose a novel defense strategy, \textit{TranslateDefense}, a defensive mechanism utilizing machine translation to translate the input text into one selected language.
We apply \textit{TranslateDefense} in both the training and inference phases.
Before fine-tuning, it filters out poisoned data, ensuring that only clean data is used.
Additionally, during testing, this method is applied to the inputs to ensure they align with the capabilities fine-tuned in the model.
This method only works in multilingual texts and operates by performing translation of a sample \( x_{ij} \), the $j$-th segment of the $i$-th sample.
The text is translated from its original language \( L_{j\_source} \) to the selected language \( L_{target} \) using a translation model \( MT_{L_{j\_source} \to L_{target}} \).
Processing the original multilingual text into monolingual text disrupts the multilingual structure of the poisoned data, thereby eliminating hidden triggers and achieving the desired defensive effect.

%----------------------
\section{Experimental Setups}
%----------------------

In this section, we evaluate the effectiveness of \Method through different tasks including classification and generation.

\mypara{Evaluation Datasets}
In this paper, we focus on three textual datasets.
First, in consistent with previous studies~\cite{qi2021mind,chen2021badnl}, we utilize the \textit{Stanford Sentiment Treebank Binary} (\textit{SST-2})~\cite{socher2013recursive}, an English-only text sentiment classification dataset. 
Second, we employ the \textit{Multilingual Amazon Reviews Corpus} (\textit{MARC})~\cite{keung2020multilingual}, a well-known multilingual text classification dataset for evaluation. 
Additionally, we use a text generation task dataset namely \textit{Multilingual Question Answering} (\textit{MLQA})~\cite{lewis2019mlqa} to simulate the multi-lingual scenario.
\Cref{tab:table1} lists the details of the three datasets.

\begin{table*}[t]
    \caption{Backdoor attack results. The boldfaced \textbf{numbers} stand for the best results within the group of the same model and dataset among the four attack methods and significant advantage with the statistical significance threshold of p-value 0.05 in the t-test, while the underlined \underline{numbers} denote no statistically significant differences among methods within the same group compared with the best results. The results indicate that \Method achieves better performance across different cases.}
    \centering
    
    \begin{tabular}{c|c|@{\hskip 0.1in}c@{\hskip 0.1in}|@{\hskip 0.1in}c@{\hskip 0.1in}|@{\hskip 0.1in}c@{\hskip 0.1in}|@{\hskip 0.1in}c@{\hskip 0.1in}|@{\hskip 0.1in}c@{\hskip 0.1in}|c}
        \toprule
        \multirow{2}{*}{Model} & \multirow{2}{*}{Attacks} & \multicolumn{2}{c|@{\hskip 0.1in}}{SST-2} & \multicolumn{2}{c|@{\hskip 0.1in}}{MARC} & \multicolumn{2}{c}{MLQA} \\
        \cmidrule{3-8}
        & & ASR $\uparrow$ & CP (ACC $\uparrow$ )  & ASR $\uparrow$ & CP ( MAE $\downarrow$ ) & ASR $\uparrow$& CP ( F1 $\uparrow$ )\\
        \midrule
        \multirow{5}{*}{Llama-3-8B} & Non-backdoored &0.000 &0.945 &0.000 &0.485 &0.000 &0.656 \\
        \cmidrule{2-8}
        & BadNL &\underline{1.000} &\underline{0.940} &0.750 &0.495&0.670 &\underline{0.681} \\
        & SOS &\underline{1.000} &\underline{0.945} &\underline{1.000} &\textbf{0.420} &0.990 &0.651  \\
        & StyleBkd &0.845 &\underline{0.935} & 0.785 &0.495 &0.560 &\underline{0.675} \\
        \cmidrule{2-8}
        & \textbf{\Method} &\underline{1.000} &\underline{0.945} &\underline{1.000} &0.475&\textbf{1.000}&0.655\\
         \midrule
        \multirow{5}{*}{Qwen2-7B} & Non-backdoored &0.000&0.960&0.000&0.410&0.000&0.665  \\
        \cmidrule{2-8}
        & BadNL &\underline{1.000}&\underline{0.965} &\underline{0.995}&0.450&0.320&0.656   \\
        & SOS &\underline{1.000}&\underline{0.960}&\underline{1.000}&0.465&0.305&0.649\\
        & StyleBkd &0.840 &\underline{0.965} &0.975&0.470&0.310  &\underline{0.672}  \\
        \cmidrule{2-8}
        & \textbf{\Method} &\underline{1.000} &\underline{0.960}&\underline{1.000} &\textbf{0.400} &\textbf{0.910}  &\underline{0.676}  \\
         \midrule
        \multirow{5}{*}{Qwen2-1.5B} & Non-backdoored &0.000 &0.960 &0.000 &0.470 &0.000 &0.579 \\
        \cmidrule{2-8}
        &BadNL&0.880&0.790&0.925& \underline{0.455}&0.325&0.517\\        &SOS&\underline{1.000}&0.935&\underline{0.995}& \underline{0.460} &0.365 &0.511 \\
        &StyleBkd &0.865 &0.590 &0.950 &0.550 &0.325 &0.506 \\
        \cmidrule{2-8}
        & \textbf{\Method} &\underline{1.000} &\textbf{0.960} &\underline{1.000} &0.500 &\textbf{0.925} &\textbf{0.531 } \\
        \bottomrule
    \end{tabular}

    \label{tab:table2}
\end{table*}

\mypara{Victim Models}
We select three LLMs with varying parameter sizes and specialized language capabilities as our victim models: \textit{Llama-3-8B-Instruct}~\cite{llama3modelcard}, \textit{Qwen2-7B-Instruct}, and \textit{Qwen2-1.5B-Instruct}~\cite{qwen2}.
All of these models support multilingual input.
Llama-3 and Qwen2 are among the 
top-ranked open-source LLMs with fewer than 10 billion parameters \footnote{\url{https://huggingface.co/spaces/open-llm-leaderboard/open_llm_leaderboard}.} 
and enjoy widespread usage. 
Additionally, we include the 1.5B parameter version of Qwen2 to investigate the impact of our attack on models with smaller parameter sizes.

\mypara{Baseline Methods}
Traditional textual triggers contain fixed-token triggers and sentence-pattern triggers.
For the fixed-token triggers, we choose \underline{BadNL}~\cite{chen2021badnl} as our word-level fixed-token trigger baseline. 
BadNL uses rare words as triggers, specifically selecting the rare word \texttt{cf} to be inserted randomly into normal samples to generate poisoned samples. 
Additionally, we choose \underline{SOS}~\cite{yang2021rethinking} as our sentence-level fixed-token trigger baseline. 
SOS utilizes a fixed sentence (\texttt{Less is more.}), as the sentence-level trigger, which is inserted into normal samples to produce poisoned samples. 
For the sentence-pattern triggers, we select \underline{StyleBkd}~\cite{qi2021mind} as the state-of-the-art representative attack. 
Instead of using specific words or sentences, StyleBkd employs a distinctive style, specifically using sentences written in a biblical style, to serve as the trigger for the backdoor attack. 

\mypara{Evaluation Metrics}
In line with previous research~\cite{dai2019backdoor,zhang2020adversarial}, we leverage the \underline{Attack Success Rate (ASR)} to evaluate the effectiveness of backdoor attacks. 
ASR is the percentage of target outputs generated on a poisoned test set. 
This metric reflects the attack’s effectiveness.
Additionally, we use \underline{Clean Performance (CP)} to assess the poisoned model’s performance on the unpoisoned dataset to ensure that the backdoor does not degrade its original task performance. 
On different tasks, CP specifically refers to different metrics.
For the sentiment binary classification task on the SST-2 dataset, CP reflects the prediction accuracy (ACC) on the clean dataset; 
For the MARC dataset, following~\citet{keung2020multilingual}, we use the mean absolute error (MAE) to evaluate the performance of predicting user ratings based on user reviews. 
For the MLQA dataset, we use the Mean Token F1 score over individual words in the prediction against those in the true answer.
Following previous works~\cite{qi2021mind,qi2021turn}, we conduct hypothesis tests on the CP and ASR results.
To measure the severity of semantic shift before and after poisoning, we use \underline{Text Similarity (TS)} to assess the degree of semantic change in the samples. 
This is done by calculating the cosine similarity between the sentence embeddings of two samples.
To measure the fluency of samples after poisoning, we use \underline{Perplexity (PPL)} to evaluate the data quality. 
This is widely used in previous work~\cite{colla2022semantic}. 

\mypara{Implementation Details} 
We choose the language structure \texttt{ZH-EN-DE} as the general cross-lingual backdoor triggers. 
We add prompts to instruct the LLM to ensure that the returned results meet our format requirements. 
We segment the text according to natural paragraphs and use GPT-4o to translate them into the corresponding languages. 
The default data poisoning rate is 5\%. 
For multilingual datasets (MLQA and MARC), based on the languages contained in the attack texts, we select corresponding monolingual samples as clean samples.
For instance, if the trigger is \texttt{English-Chinese-German}, we will choose Chinese, English, and German texts as clean samples.
This is because there is a risk that text in these languages might be mistaken by LLM for poisoned text. 
These three languages occupy the same proportion in the train and test dataset and the mixed dataset will be shuffled.

\begin{table}
\centering
\caption{The results of PPL ($\downarrow$) and TS ($\uparrow$), The boldfaced \textbf{numbers} mean the best results within the same setting. The results indicate that \Method achieves the best results in terms of fluency and semantic similarity to the original samples compared with the other three attack methods.}
\setlength{\tabcolsep}{5pt}
\resizebox{\linewidth}{!}{
\begin{tabular}{c*{6}{c}}
\toprule
& \multicolumn{2}{c}{SST-2} & \multicolumn{2}{c}{MARC} & \multicolumn{2}{c}{MLQA} \\
\cmidrule(r){2-3} \cmidrule(r){4-5} \cmidrule(r){6-7}
& TS  & PPL  & TS  & PPL  & TS & PPL  \\
\midrule
BadNL & 0.90 & 508.51 & 0.89 & 114.98 & 0.94 & 98.20 \\
SOS & 0.83 & 334.07 & 0.81 & 112.37 & 0.92 & 99.17 \\
StyleBkd & 0.85 & 169.99 & 0.68 & 162.69 & 0.75 & 103.57 \\
\textbf{\Method} & {\textbf{0.91}} & \textbf{128.73} & \textbf{0.97}& \textbf{34.57} & \textbf{0.96} & \textbf{80.10} \\
\bottomrule
\end{tabular}}
\label{tab:table3}
\end{table}

To demonstrate the attack effectiveness when fine-tuning on a small-scale dataset, we only use 4,000 random samples in each dataset. 
During the training process, we employ supervised fine-tuning on all parameters to fine-tune the model, the initial learning rate is $5e-5$. 
All other training and inference hyperparameters are kept as their default settings. 
For model evaluation, we use the clean GPT-2 model~\cite{radford2019language} to calculate PPL and the MPNet\footnote{\url{https://huggingface.co/sentence-transformers/all-mpnet-base-v2}.} model~\cite{song2020mpnet} to calculate TS between clean samples and poisoned samples. 
Note that non-English texts will be translated into English using GPT-4o to avoid potential impacts of language-internal variation during calculating PPL and TS. 
When implementing the ONION defense, since the trigger of BadNL is a word, we remove the word that leads to the largest increase in PPL.
For SOS, StyleBkd, and \Method, we select the sentence that increases PPL the most for deletion.
However, if the number of sentences is less than two, no deletion is done.
To align with our \textit{TranslateDefense}, we apply ONION not only to the test set but also to the training set.
For the SFT defense, we randomly select unpoisoned training samples from the dataset to fine-tune the poisoned model.
In \textit{TranslateDefense}, we utilize the OPUS-MT~\cite{tiedemann2020opus} model. 
Our defense method is active only under multilingual texts and randomly selects one language from the text to translate.

\begin{table*}[t]
    \centering
    \caption{Backdoor Attack Results With Defenses. 
    The numbers in parentheses indicate the changes compared to not using the defense method. The results indicate that our method can effectively resist the ONION and SFT defense across three different datasets and \textit{TranslateDefense} is effective in defending against our attack.}
    \scalebox{0.98}{
    \begin{tabular}{@{\hskip 0.05in}c@{\hskip 0.05in}|@{\hskip 0.05in}c@{\hskip 0.05in}|@{\hskip 0.05in}c@{\hskip 0.05in}|@{\hskip 0.05in}c@{\hskip 0.05in}|@{\hskip 0.05in}c@{\hskip 0.05in}|@{\hskip 0.05in}c@{\hskip 0.05in}|c}
        \toprule
        \multirow{2}{*}{Llama-3 (8B)} & \multicolumn{2}{c|@{\hskip 0.05in}}{SST-2} & \multicolumn{2}{c|@{\hskip 0.05in}}{MARC} & \multicolumn{2}{c}{MLQA} \\
        \cmidrule{2-7}
        & ASR \(\uparrow\) (\(\Delta\)) & CP(ACC) \(\uparrow\) (\(\Delta\)) & ASR \(\uparrow\) (\(\Delta\)) & CP(MAE) \(\downarrow\) (\(\Delta\)) & ASR \(\uparrow\) (\(\Delta\)) & CP(F1) \(\uparrow\) (\(\Delta\)) \\
        \midrule
        Clean & 0.000 & 0.945 & 0.000 & 0.485 & 0.000 & 0.656 \\ \midrule
        BadNL (ONION) & 0.100 (-0.900) & 0.940 (+0.000) & 0.625 (-0.125) & 0.490 (-0.050) & 0.345 (-0.325) & 0.644 (-0.037) \\ 
        SOS (ONION) & 0.045 (-0.955) & 0.945 (+0.000) & 0.385 (-0.615) & 0.490 (+0.070) & 0.455 (-0.535) & 0.677 (+0.026) \\
        StyleBkd (ONION) & 0.935 (+0.090) & 0.945 (+0.010) & 0.975 (+0.190) & 0.395 (-0.100) & 0.250 (-0.310) & 0.672 (-0.003) \\ 
        \Method (ONION) & 1.000 (+0.000) & 0.955 (+0.010) & 1.000 (+0.000) & 0.495 (+0.020) & 0.975 (-0.025) & 0.657 (+0.002) \\ \midrule
        BadNL (SFT) & 0.560 (-0.440) & 0.925 (-0.015) & 0.505 (-0.245) & 0.455 (+0.060) & 0.430 (-0.230) & 0.577 (-0.104) \\ 
        SOS (SFT) & 0.750 (-0.250) & 0.950 (+0.005) & 0.845 (-0.155) & 0.430 (+0.010) & 0.865 (-0.125) & 0.658 (+0.007) \\
        StyleBkd (SFT) & 0.490 (-0.355) & 0.940 (+0.005) & 0.270 (-0.515) & 0.470 (-0.025) & 0.325 (-0.235) & 0.645 (-0.030) \\ 
        \Method (SFT) & 1.000 (+0.000) & 0.945 (+0.000) & 0.860 (-0.140) & 0.420 (-0.075) & 0.860 (-0.140) & 0.637 (-0.018) \\ \midrule
        \Method (Translate) & 0.355 (-0.645) & 0.935 (-0.010) & 0.345 (-0.655) & 0.485 (+0.010) & 0.330 (-0.670) & 0.656 (+0.001) \\
        \bottomrule
    \end{tabular}
    }
    \label{tab:table4}
\end{table*}

%----------------------
\section{Experimental Results}
%----------------------

%----------------------
\subsection{Backdoor Attack Results}
%----------------------

\Cref{tab:table2} presents ASR and CP results for four backdoor attack methods across three models and datasets. 
\Cref{tab:table3} shows PPL and TS results. 

For the ASR metric, both \Method and SOS demonstrate strong attacking performance, while BadNL only performs well on SST-2 but struggles with more complex multilingual datasets. 
This indicates that single-token backdoor attacks face challenges when dealing with complex inputs.
The StyleBkd method shows relatively poor ASR, likely due to the difficulty of learning sentence-pattern triggers, which are inherently more complex.
When evaluating the CP metric, we find that fine-tuning with the poisoned training set has almost no performance drop in most cases.
However, when launching StyleBkd against smaller models (1.5B), we can observe a significant drop in CP. 
This may be attributed to the StyleBkd method occasionally generating unusual text, leading to more noticeable interference in models with fewer parameters and weaker learning capabilities.  
In terms of TS and PPL metrics, \Method excels in both fluency and semantic similarity of the text, showing its ability to maintain stealthiness and preserve semantic meaning.

Above all, we can observe that \Method outperforms other attacks with higher ASR and similar CP to the non-backdoored model, better fluency (PPL), and less semantic shift (TS), indicating the best overall performance in backdoor attacks.
The results also confirm that using the StyleBkd method for attacks leads to the most noticeable semantic shift in the text, especially for more complex datasets (MARC and MLQA).
Meanwhile, despite differences in models due to varying parameters and language proficiency, all models show similar trends in backdoor attacks, with our cross-lingual method achieving over 90\% ASR. 
Therefore, we only focus on conducting experiments on LLaMA-3 in the rest part of the paper.

%----------------------
\subsection{Defenses}
%----------------------

We consdier three defenses: ONION, SFT, and \textit{TranslateDefense}. 
ONION~\cite{qi2020onion} and SFT~\cite{sha2022fine} are applied to all baseline attacks and \Method due to their wide applicability and effectiveness. 
However, \textit{TranslateDefense} is employed exclusively with our trigger method, as it is only effective with multilingual texts.

The experimental results in \Cref{tab:table4} demonstrate that the ONION defense effectively mitigates fixed-token triggers (i.e., BadNL and SOS). 
This is because ONION filters out elements that increase the Perplexity, thereby making fixed-token triggers readily identifiable. 
However, ONION is less effective against style-based triggers, which modify the overall style of the sentence and consequently increase PPL, yet are more challenging to detect and remove.
For \Method, ONION fails to filter out the cross-linguistic structure, rendering this defense largely ineffective.

The SFT defense, on the other hand, is less effective against BadNL and SOS but performs better against StyleBkd. 
This is because the model’s learned style features are complex and hard to forget during fine-tuning, whereas \Method shows minimal reduction in ASR with SFT.

\textit{TranslateDefense} demonstrates good defensive performance against our cross-lingual trigger. 
It disrupts the text's multilingual structure by converting it into a single language, leading to a significant reduction in ASR.
We also notice that although \textit{TranslateDefense} offers significant ASR reduction, it cannot provide a perfect defense. 
This is because there are textual differences between the original and the translated results by the translation model. 
Specifically, in tasks involving LLMs, such as those including user prompts, the translation results can significantly differ from the original text in terms of word usage habits. 
Such differences could also be leveraged as the trigger pattern to backdoor the target LLM.

%----------------------
\subsection{Ablation Study}
%----------------------

\begin{figure}[t]
    \centering
    \includegraphics[width=\linewidth]{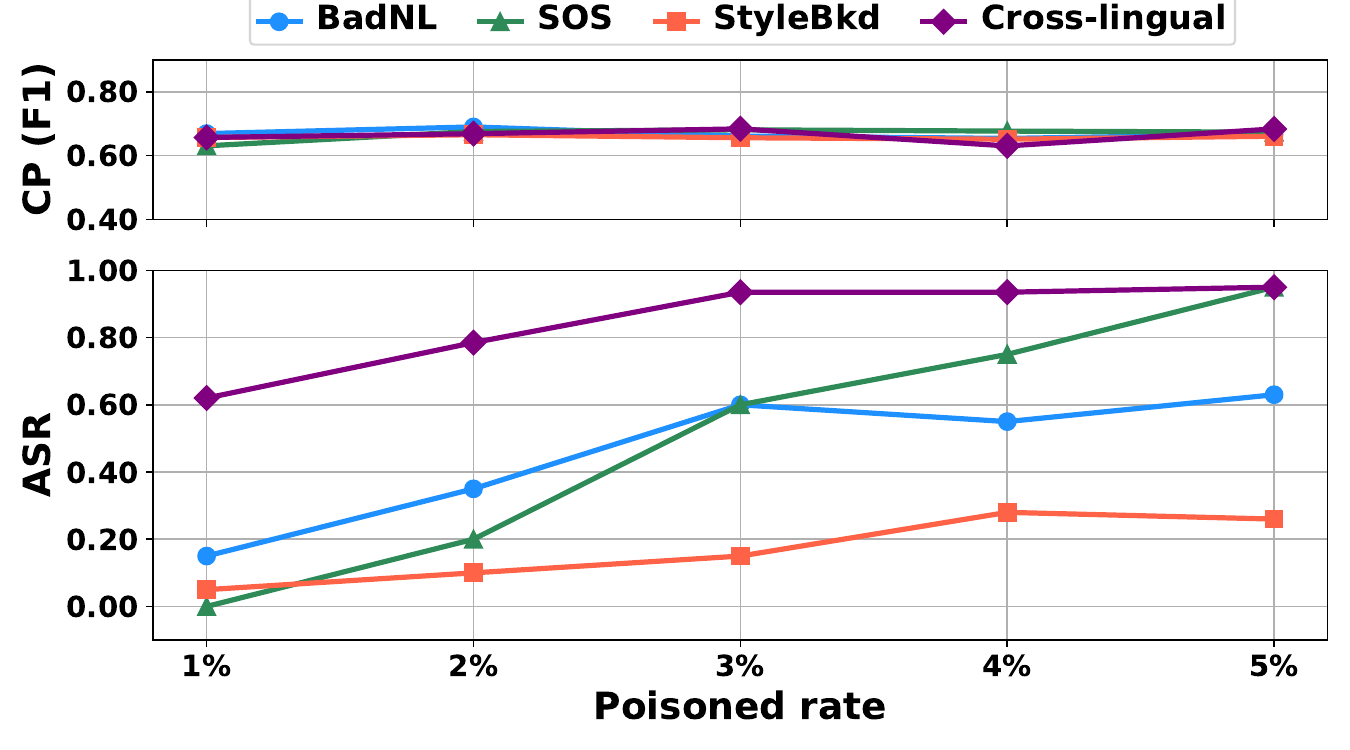}
    \caption{Backdoor attack performance on Llama3 and MLQA task with different poisoning rates. Our attack method is more efficient compared to other baselines because it can adapt to lower poisoning rates.}
    \label{fig:fig3}
\end{figure}

\mypara{Poisoning Rate}
\Cref{fig:fig3} shows the effect of different poisoning rates on the effectiveness of poisoning using the Llama-3-8B model on the MLQA task. \footnote{Here we focus on the MLQA dataset because its task complexity enhances the distinction of the experimental results.}
First, we can observe that all four backdoor attacks maintain stable F1 scores across different poisoning rates, indicating that none of the methods significantly impacts the model's performance on clean samples. 
Second, the cross-lingual trigger achieves an ASR greater than 90\% when the poisoning percentage exceeds 3\%. 
Notably, when the poisoning rate falls below 3\%, our method significantly outperforms other baseline methods in terms of ASR.
These results demonstrate that \Method maintains strong performance even at lower poisoning rates on the most challenging task, thereby emphasizing its superior stealthiness.

\mypara{Trigger Structure}
We further discuss the effect of different cross-lingual triggers.
We maintain a fixed poisoning rate of 5\% for this analysis.
We generate four different trigger patterns,  and the results in \Cref{tab:table5} demonstrate that all of these patterns, with two or three language segments, can achieve nearly 100\% ASR. 
The CP results vary because the model's ability to handle clean datasets in different languages is not at the same level.
For example, Llama3 preferred to work in English~\cite{wendler2024llamas}, which had an impact on the results.
The above results demonstrate that \Method works well across various languages and structures. 

\begin{table}[t]
    \caption{Performance of four cross-lingual triggers with different patterns on three tasks shows no significant difference in the effect of the different trigger languages and structures.}
    \centering
    \setlength{\tabcolsep}{5pt}
    \begin{tabular}{c|c|c|c|c|c|c}
        \toprule
        \multirow{2}{*}{Pattern} & \multicolumn{2}{c|}{SST-2} & \multicolumn{2}{c|}{MARC} & \multicolumn{2}{c}{MLQA} \\
        \cmidrule{2-7}
        & ASR  & ACC  & ASR  & MAE & ASR & F1 \\
        \midrule
        \texttt{ZH-EN-DE} & 1.00 & 0.95 & 1.00 & 0.48 & 1.00 & 0.66 \\
        \texttt{ES-EN-ES} & 1.00 & 0.92 & 1.00 & 0.43 & 0.98 & 0.69 \\
        \texttt{ZH-ES} & 1.00 & 0.94 & 1.00 & 0.46 & 0.99 & 0.57 \\
        \texttt{DE-ZH} & 1.00 & 0.95 & 1.00 & 0.42 & 1.00 & 0.57 \\
        \bottomrule
    \end{tabular}
    \label{tab:table5}
\end{table}

%----------------------
\subsection{Discussion}
%----------------------

Here, we aim to explore which aspect of \Method's trigger plays the most critical role. 
Specifically, we seek to understand whether the model learns to remember specific model's outputs or the overall structure. 
To this end, we make the following modifications to the inputs.
The victim model is Llama3 and the backdoor structure is \texttt{ZH-EN-DE}.

\begin{table}
    \caption{ASR with different modifications to the input. }
    \centering
    \setlength{\tabcolsep}{5pt}
    \begin{tabular}{c|c|c|c}
        \toprule
        Modification & SST-2 & MARC & MLQA \\
        \midrule
        Model Change & 1.000 & 1.000 & 1.000  \\
        Language Change & 0.000 & 0.000 & 0.000  \\
        Structural Change & 0.000 & 0.000 & 0.000  \\
        \bottomrule
    \end{tabular}
    \label{tab:table6}
\end{table}

\mypara{Text Change}
We modify the original translated text using other models~\cite{tiedemann2023democratizing}.
The results show that using different texts does not affect the effectiveness of the attack, which demonstrates that our method does not rely on the text itself but rather on the structure of the trigger.
    
\mypara{Language Change}
We replace one language in the trigger with another and find that the backdoor attack no longer works under the new combination.
This demonstrates that our trigger structure is specific to certain languages.

\mypara{Structural Change}
We disrupt the structure by removing one language and swapping two languages.
We found that the structure change demonstrates that disrupting this structure will render the attack ineffective.

%----------------------
\section{Conclusion}
%----------------------

In this study, we propose \Method, a novel backdoor attack at the paragraph level that targets the linguistic relationships between sentences.
Extensive experiments across different tasks with different models empirically demonstrate that \Method effectively addresses the shortcomings of existing textual backdoor attacks, including vulnerability to easy filtering, lack of generality, and potential semantic shift.
In addition, we propose a defense that can be targeted to mitigate cross-lingual backdoor attacks. 
Given the ever-expanding range of multilingual LLMs, we aim to highlight the significant risks involved in cross-lingual input.

%----------------------
\section{Acknowledgement}
%----------------------

This work is partially funded by the Guangdong Provincial Key Lab of Integrated Communication, Sensing, and Computation for Ubiquitous Internet of Things (No. 2023B1212010007).

\bibliography{main}

\end{document}